\documentclass[12pt]{article}
\usepackage{amsmath}
\usepackage{graphicx}
\usepackage{bm}
\usepackage{fancyhdr}
\usepackage{amssymb}
\usepackage{setspace}
\usepackage{epsfig}
\usepackage{overpic}
\usepackage{epstopdf}
\usepackage{footmisc}
\usepackage{caption,subcaption}
\allowdisplaybreaks[4]
\renewcommand{\Re}{\operatorname{Re}}
\renewcommand{\Im}{\operatorname{Im}}

\begin{document}

\baselineskip=20pt


\newcommand{\Title}[1]{{\baselineskip=26pt   \begin{center} \Large \bf #1 \\ \ \\ \end{center}}}
\newcommand{\Author}{\begin{center}    \large \bf
Xiong Le${}^{a,b}$, Yi Qiao${}^{a,b,}\footnote{Corresponding author.\label{Corr}}$, Junpeng Cao${}^{a,b,c,d,}\footref{Corr}$, Wen-Li Yang${}^{d,e,f,g,}\footref{Corr}$, Kangjie Shi${}^{e}$, and Yupeng Wang${}^{a,d,h}$
\end{center}}

\newcommand{\Address}{ \begin{center}
     ${}^a$ Beijing National Laboratory for Condensed Matter Physics, Institute of Physics, Chinese Academy of Sciences, Beijing 100190, China\\
     ${}^b$ School of Physical Sciences, University of Chinese Academy of Sciences, Beijing 100049, China\\
     ${}^c$ Songshan Lake Materials Laboratory, Dongguan, Guangdong 523808, China\\
     ${}^d$ Peng Huanwu Center for Fundamental Theory, Xian 710127, China\\
     ${}^e$ Institute of Modern Physics, Northwest University, Xian 710127, China\\
     ${}^f$ School of Physics, Northwest University, Xian 710127, China\\
     ${}^g$ Shaanxi Key Laboratory for Theoretical Physics Frontiers, Xian 710127, China\\
     ${}^h$ The Yangtze River Delta Physics Research Center, Liyang, Jiangsu, China
   \end{center}}

\newcommand{\Email}{ \begin{center} E-mail: lexiong@iphy.ac.cn, qiaoyi\_joy@foxmail.com, junpengcao@iphy.ac.cn, wlyang@nwu.edu.cn, kjshi@nwu.edu.cn, yupeng@iphy.ac.cn
\end{center}}

\Title{Root patterns and energy spectra of quantum integrable systems without $U(1)$ symmetry: the antiperiodic $XXZ$ spin chain}
\Author

\Address

\Email

\vspace{0.2truecm}

\begin{abstract}

Finding out root patterns of quantum integrable models is an important step to study their physical properties in the thermodynamic limit. Especially for models without $U(1)$ symmetry, their spectra are usually given by inhomogeneous $T-Q$ relations and the Bethe root patterns are still unclear. In this paper with the antiperiodic $XXZ$ spin chain as an example, an analytic method to derive both the Bethe root patterns and the transfer-matrix root patterns in the thermodynamic limit is proposed. Based on them the ground state energy and elementary excitations in the gapped regime are derived. The present method provides an universal procedure to compute physical properties of quantum integrable models in the thermodynamic limit.

\vspace{1truecm}

\noindent {\it PACS:} 75.10.Pq, 03.65.Vf, 71.10.Pm\\
\noindent {\it Keywords}: Root patterns; Inhomogeneous parameters; Antiperiodic $XXZ$ chain;
\end{abstract}

\newpage

\section{Introduction}
\setcounter{equation}{0}

The exactly solvable models play important roles in modern physics and mathematics. These models can provide crucial benchmarks for important physical concepts and phenomena such as thermodynamic phase transitions from the two-dimensional Ising model \cite{C1-4}, the Mott insulator from the one-dimensional Hubbard model \cite{C1-5}, fractional charges from the Heisenberg spin chain \cite{C1-6} and etc. In the past decades, several methods including the coordinate Bethe Ansatz \cite{C1-13}, the $T - Q$ relation \cite{C1-14,C1-15} and the algebraic Bethe Ansatz \cite{C1-16,C1-17,C1-18,C1-19,C1-20,C1-21,C1-22} were developed. These methods work quite well for models with obvious reference states because the root patterns are clear \cite{Takahashi99}. For the quantum integrable systems without $U(1)$ symmetry, which have important applications in non-equilibrium statistical physics \cite{Q3-6,Q3-7}, condensed matter physics \cite{Q3-5} and high energy physics \cite{Q3-1}, their spectra are usually described by an inhomogeneous $T-Q$ relation \cite{Cao13-1,Book}. The inhomogeneous term in the Bethe ansatz equations (BAEs) makes the problem complicated since the Bethe root patterns are not clear. Therefore, finding out root patterns is an important step to compute physical properties of corresponding systems. Several authors had made important conjectures for the Bethe root patterns of some models \cite{nepo13,Xin18,Qiao18,Qiao20,Qiao21} based on numerical simulations for finite size systems.

In this paper, with the antiperiodic $XXZ$ spin chain as a concrete example, we propose an analytic method to derive both Bethe root patterns and transfer-matrix root patterns of quantum integrable models without $U(1)$ symmetry. The paper is organized as follows. Section \ref{2} serves as an introduction to the antiperiodic $XXZ$ spin chain, a typical quantum integrable model without $U(1)$ symmetry.
In section \ref{3}, we show the root patterns of the eigenvalue of the transfer matrix.
In section \ref{4}, we compute the ground state energy and the elementary excitations based on the root patterns for $\eta \in \mathbb{R}$ (ferromagnetic regime).
Section \ref{5} is attributed to the case of $\eta \in \mathbb{R}+i\pi$ (anti-ferromagnetic regime).
Concluding remarks are given in section \ref{6}.

\section{Antiperiodic XXZ spin chain}\label{2}
\setcounter{equation}{0}

The Hamiltonian of the antiperiodic $XXZ$ spin chain \cite{batchelor95} reads
\begin{eqnarray}\label{Ham}
H=-\sum_{j=1}^N\big[ \sigma_j^x \sigma_{j+1}^x + \sigma_j^y \sigma_{j+1}^y +\cosh\eta \sigma_j^z \sigma_{j+1}^z \big],
\end{eqnarray}
where $N$ is the number of sites, $\sigma_j^\alpha (\alpha=x, y, z)$ are the Pauli matrices on $j$th site and $\eta$ is the anisotropic or crossing parameter.
We consider $\eta \in \mathbb{R}$ and $\eta \in {\mathbb{R}+i\pi}$, corresponding to the ferromagnetic regime and the anti-ferromagnetic regime, respectively. The antiperiodic boundary condition is achieved by
\begin{eqnarray}\label{Anti-periodic}
\sigma_{N+1}^\alpha=\sigma_1^x\sigma_1^\alpha\sigma_1^x,\quad {\rm for}\quad \alpha=x, y, z,
\end{eqnarray}
which breaks the $U(1)$-symmetry of the system.

The integrability of the model (\ref{Ham}) is associated with the six-vertex $R$-matrix
\begin{eqnarray}\label{R-matrix}
	R_{0,j}(u)= \frac{1}{2} \left[ \frac{\sinh(u+\eta)}{\sinh \eta} (1+\sigma^z_j \sigma^z_0) +\frac{\sinh u}{\sinh \eta} (1- \sigma^z_j \sigma^z_0)   \right]+ \frac{1}{2} (\sigma^x_j \sigma^x_0 +\sigma^y_j \sigma^y_0),
\end{eqnarray}
where $u$ is the spectral parameter. The $R$-matrix is defined in the auxiliary space $V_0$ and the quantum space $V_j$ and satisfies
\begin{eqnarray}\label{R-matrix relations}
&&\hspace{-0.5cm}\mbox{ Initial
	condition}:\,R_{0,j}(0)=P_{0,j},\nonumber \\
&&\hspace{-0.5cm}\mbox{ Unitary
	relation}:\,R_{0,j}(u)R_{j,0}(-u)=\phi(u)\times {\rm id},\nonumber \\
&&\hspace{-0.5cm}\mbox{ Crossing
	relation}:\,R_{0,j}(u)=-\sigma_0^y R_{0,j}^{t_0}(-u-\eta)\sigma_0^y,\nonumber \\
&&\hspace{-0.5cm}\mbox{ PT-symmetry}:\,R_{0,j}(u)=R_{j,0}(u)=R_{0,j}^{t_0\,t_j}(u),
\end{eqnarray}
where $P_{0,j}$ is the permutation operator,
$\phi(u)=-\sinh(u+\eta)\sinh(u-\eta)/\sinh^2\eta$, $t_0$ means the transposition in the auxiliary space and $t_j$ means the transposition in the $j$th space. Besides,
the $R$-matrix (\ref{R-matrix}) also satisfies the Yang-Baxter equation
\begin{eqnarray}\label{Yang-Baxter equation}
R_{1,2}(u_1-u_2)R_{1,3}(u_1-u_3)R_{2,3}(u_2-u_3)
=R_{2,3}(u_2-u_3)R_{1,3}(u_1-u_3)R_{1,2}(u_1-u_2).
\end{eqnarray}
The transfer matrix $t(u)$ of the system is constructed by the $R$-matrix (\ref{R-matrix}) as
\begin{eqnarray}
t(u)=tr_0 \{ \sigma_0^x R_{0,N}(u-\theta_N) \cdots R_{0,1}(u-\theta_1) \},\label{transfer_matrix}
\end{eqnarray}
where $\{\theta_j|j=1, \cdots, N\}$ are the site-dependent inhomogeneity parameters and $tr_0$ means the partial trace over the auxiliary space. From the Yang-Baxter equation (\ref{Yang-Baxter equation}), one can prove that the transfer matrices with different spectral parameters commutate with each other, i.e., $[t(u),t(v)]=0$. Therefore, the transfer matrix $t(u)$ is the generating function of all the conserved quantities of the system. The model Hamiltonian (\ref{Ham}) is related to the transfer matrix as
\begin{eqnarray}\label{def-Ham}
	H=-2 \sinh \eta  \frac{\partial \ln t(u)}{\partial u}\big|_{u=0, \{\theta_j=0\}} + N \cosh \eta.
\end{eqnarray}

Using the properties of the $R$-matrix (\ref{R-matrix relations}), we obtain the following operator identities \cite{Book}
\begin{eqnarray}\label{tt_relation}
	t(\theta_j)t(\theta_j-\eta)=-a(\theta_j)d(\theta_j-\eta)\times id, \quad j=1, \cdots N,
\end{eqnarray}
where
\begin{eqnarray}\label{ad_func}
d(u)=a(u-\eta)=\prod_{j=1}^N \frac{\sinh(u-\theta_j)}{\sinh \eta}.
\end{eqnarray}
From the definition (\ref{transfer_matrix}), we know that the transfer matrix $t(u)$ is a trigonometrical polynomial operator of $u$ with the degree $N-1$.
Besides, it satisfies the periodicity $t(u+i\pi)=(-1)^{N-1}t(u)$.
The transfer matrix $t(u)$ can be rewritten as
\begin{eqnarray}\label{ai1}
t(u)=(-1)^{N-1}tr_0 \{\sigma^x_0R_{0,N}^{t_{0}}(-u-\eta)R_{0,N-1}^{t_{0}}(-u-\eta) \cdots R_{0,1}^{t_{0}}(-u-\eta)\}.
\end{eqnarray}
If $\eta \in \mathbb{R}$ or $\eta \in {\mathbb{R}+i\pi}$, the $R$-matrix satisfies the relation
\begin{eqnarray}\label{R-R-relation}
	R_{0,j}^{\ast t_{j}}(-u-\eta)=R_{0,j}^{t_{0}}(-u^{\ast}-\eta).
\end{eqnarray}
Substituting Eq.(\ref{R-R-relation}) into Eq.(\ref{ai1}) and taking the Hermitian conjugate, we obtain
\begin{eqnarray}\label{t-t-relation}
t^{\dagger}(u)=(-1)^{N-1}t(-u^{\ast}-\eta).
\end{eqnarray}

Denote the eigenvalue of the transfer matrix $t(u)$ as $\Lambda(u)$. From above analysis, we know that the
eigenvalue $\Lambda(u)$ satisfies
\begin{eqnarray}
&&	\Lambda(\theta_j)\Lambda(\theta_j-\eta)=-a(\theta_j)d(\theta_j-\eta), \quad j=1, \cdots N,\label{TT_relation} \\
&& \Lambda(u+i\pi)=(-1)^{N-1}\Lambda(u),  \label{ai2}\\
&&\Lambda(u)=(-1)^{N-1}\Lambda^{\ast}(-u^{\ast}-\eta).\label{ai3}
\end{eqnarray}
Obviously, $\Lambda(u)$ is a degree $N-1$ trigonometric polynomial of $u$ and can be parameterized as
\begin{eqnarray}\label{lam}
	\Lambda(u)=\Lambda_0\prod_{j=1}^{N-1}\sinh(u-z_j+\frac{\eta}{2}),
\end{eqnarray}
where  $\Lambda_0$ is a coefficient and $\{z_j|j=1, \cdots, N-1\}$ are the zero roots of the polynomial.
The constraints (\ref{TT_relation}) determine the $N$ unknowns $\Lambda_0$ and $\{z_j| j=1, \cdots, N-1\}$ completely.
The energy spectrum of the Hamiltonian (\ref{Ham}) is determined by the zero roots $\{z_j\}$ as
\begin{eqnarray}\label{energy}
	E=2\sinh\eta\sum_{j=1}^{N-1}\coth(z_j-\frac{\eta}{2})+N\cosh\eta.
\end{eqnarray}

\section{Patterns of zero roots}\label{3}
\setcounter{equation}{0}

From (\ref{ai3}) we deduce that for any given root $z_j$, there must be another root $z_l$ satisfy
\begin{eqnarray}\label{tanqi}
z_j+z^{\ast}_l=k i\pi, \quad k\in \mathbb{Z}.
\end{eqnarray}
Therefore, with the periodicity Eq.(\ref{ai2}) and Eq.(\ref{tanqi}), we find that the zero roots $\{z_j\}$ can be classified into two types
\begin{eqnarray}
&& {\rm (i)}  \;\; \; \Re({z_j})=0, \quad \Im({z_j})\in \left[-\frac{\pi}{2},\frac{\pi}{2}\right), \label{tanqi1} \\
&& {\rm (ii)} \;\; \Re({z_j})+\Re({z_l})=0, \quad \Im({z_j})=\Im({z_l})\in \left[-\frac{\pi}{2},\frac{\pi}{2}\right).\label{tanqi2}
\end{eqnarray}
In case (i), all the zero roots are on the imaginary axis. Then we should analyze the patterns of zero roots in case (ii).

Let us consider first the positive $\eta$ case.
According to the functional relations (\ref{TT_relation}), the eigenvalue $\Lambda(u)$ can also be expressed as the inhomogeneous $T - Q$ relation \cite{Book}
\begin{eqnarray}\label{T-Q relation}
		\Lambda(u)=e^{u}a(u)\frac{Q(u-\eta)}{Q(u)}-e^{-u-\eta}d(u)\frac{Q(u+\eta)}{Q(u)}-c(u)\frac{a(u)d(u)}{Q(u)},
\end{eqnarray}
where the functions $Q(u)$ and $c(u)$ are given by
\begin{eqnarray}\label{Q}
&&Q(u)=\prod_{j=1}^{N}\frac{\sinh(u-\lambda_j)}{\sinh\eta}, \nonumber \\
&&c(u)=e^{u-N\eta+\sum_{l=1}^{N}(\theta_{l}-\lambda_l)}-e^{-u-\eta-\sum_{l=1}^{N}(\theta_{l}-\lambda_l)},
\end{eqnarray}
and $\{\lambda_j\}$ are the Bethe roots. Putting $\lambda_j\equiv iu_j-\frac{\eta}{2}$ and considering the homogeneous limit $\{\theta_j\rightarrow0\}$,
the Bethe roots $\{u_j\}$ should satisfy the BAEs
\begin{eqnarray}\label{associated BAEs Conc}
&&e^{iu_j}\left[\frac{\sin(u_j-\frac{1}{2}i\eta)}{\sin(u_j+\frac{1}{2}i\eta)}\right]^{N}=e^{-iu_j}\prod_{l=1}^{N}\frac{\sin(u_j-u_l-i\eta)}{\sin(u_j-u_l+i\eta)}\nonumber \\
&&+2ie^{-\frac{1}{2}N\eta}\sin(u_j-\sum_{l=1}^{N}u_{l})\prod_{l=1}^{N}\frac{\sin(u_j-\frac{1}{2}i\eta)}{\sin(u_j-u_l+i\eta)}, \quad j=1,\cdots,N.
\end{eqnarray}
For a complex Bethe root $u_j$ with a negative imaginary part, we readily have
\begin{eqnarray}\label{negative u part}
\left|\sin(u_j-\frac{1}{2}i\eta)\right|> \left|\sin(u_j+\frac{1}{2}i\eta)\right|.
\end{eqnarray}
This indicates that the module of the left hand side of Eq.(\ref{associated BAEs Conc}) is larger than 1. Thus in the thermodynamic limit $N\to\infty$, the
left hand side tends to infinity exponentially. To keep Eq.(\ref{associated BAEs Conc}) holding, the right hand side of Eq.(\ref{associated BAEs Conc}) must also tends to infinity in the same order.
We note that the last term in the inhomogeneous BAEs (\ref{associated BAEs Conc}) tends to zero due to the existence of factor $e^{-\frac{1}{2}N\eta}$ with $N\rightarrow\infty$, which can be neglected in the thermodynamic limit. Thus the denominator of the first term in the right hand side must tend to zero exponentially, which leads to $u_j-u_l+i\eta\rightarrow0$.
From Eq.(\ref{T-Q relation}), we learn that the zero roots of $\Lambda(u)Q(u)$ are $z_j-\frac{\eta}{2}$ and $iu_j-\frac{\eta}{2}$, which are undistinguishable, so $u_j$ are symmetric about the real axis from the fact that $z_j$ are symmetric about the imaginary axis.
Then the Bethe roots form strings
\begin{eqnarray}\label{u_j}
u_j=u_{j0}+i\eta\Big(\frac{n+1}{2}-j\Big)+o(e^{-\delta N}),\quad j=1,\cdots,n, \quad n=1,2,\cdots,
\end{eqnarray}
where $u_{j0}$ is the position of the string in the real axis, $n$ is the length of string and $o(e^{-\delta N})$ denotes the infinitesimal correction.
If $n=1$, the Bethe root is real.

The structure of zero roots of eigenvalue $\Lambda(u)$ can be obtained by Eq.(\ref{T-Q relation}). Substituting the zero roots $\{z_j\}$ into Eq.(\ref{T-Q relation}), we obtain the relation between $u_j$ and $z_j$.
\begin{eqnarray}\label{associated BAEs Conc u z}
&&e^{iz^{\prime}_j-\frac{\eta}{2}}\left[\frac{\sin(z^{\prime}_j-\frac{i\eta}{2})}{\sin(z^{\prime}_j+\frac{i\eta}{2})}\right]^{N}
=e^{-iz^{\prime}_j-\frac{\eta}{2}}\prod_{l=1}^{N}\frac{\sin(z^{\prime}_j-u_l-i\eta)}{\sin(z^{\prime}_j-u_l+i\eta)}\nonumber \\
&&+c(iz^{\prime}_j-\frac{\eta}{2})\frac{\sin^{N}(z^{\prime}_j-\frac{i\eta}{2})}{\prod_{l=1}^{N}\sin(z^{\prime}_j-u_l+i\eta)}, \quad j=1,\cdots,N,
\end{eqnarray}
where $z_j\equiv iz^{\prime}_j$.
The rest analysis is similar with before. If $z^{\prime}_j$ has a negative imaginary part, the left hand side of Eq.(\ref{associated BAEs Conc u z}) will tend to infinity with $N$ tends to infinity.
Because the function $c(iz^{\prime}_j-\frac{\eta}{2})$ in Eq.(\ref{associated BAEs Conc u z}) also tends to 0 if $N\rightarrow \infty$, we neglect the third term in Eq.(\ref{associated BAEs Conc u z}).
The validity of Eq.(\ref{associated BAEs Conc u z}) requires that the denominator of the first term in the right hand side should tend to zero, which leads to one zero root and one Bethe root satisfying
$z^{\prime}_j-u_l+i\eta \to 0$. We should note the roots of functions $\Lambda(u)$ and $Q(u)$ could not be equal, i.e., $z^{\prime}_j\neq u_i$. Thus from the structure of Bethe roots $\{u_l\}$, we obtain the structure of $\{z^{\prime}_j\}$ as
\begin{eqnarray}
\Im(z^{\prime}_j)=-{\eta}\frac{n+1}{2}+o(e^{-\delta N}),\quad n=1,2,3,\cdots,
\end{eqnarray}
which are in the lower complex plane.
Because the zero roots $\{z^\prime_j\}$ are symmetric about the real axis, we arrive at
\begin{eqnarray}\label{zdis}
		\Re({z_j})=\pm\frac{(n+1){\eta}}{2}+o(e^{-\delta N}),\quad n=1,2,3,\cdots.\label{zwdis}
\end{eqnarray}
Therefore, we conclude that the zero roots $z_j$ are either imaginary or anti-conjugate pairs given by (\ref{zwdis}). This conclusion
is also hold for  $\eta \in \mathbb{R}+i\pi$ by changing $\eta$ to $\Re(\eta)$.

\section{Exact solution for $\eta \in \mathbb{R}$ }\label{4}
\setcounter{equation}{0}

\begin{figure*}[t]
	\begin{minipage}[b]{0.40\textwidth}
		\centering
		\includegraphics[height=6cm,width=8cm]{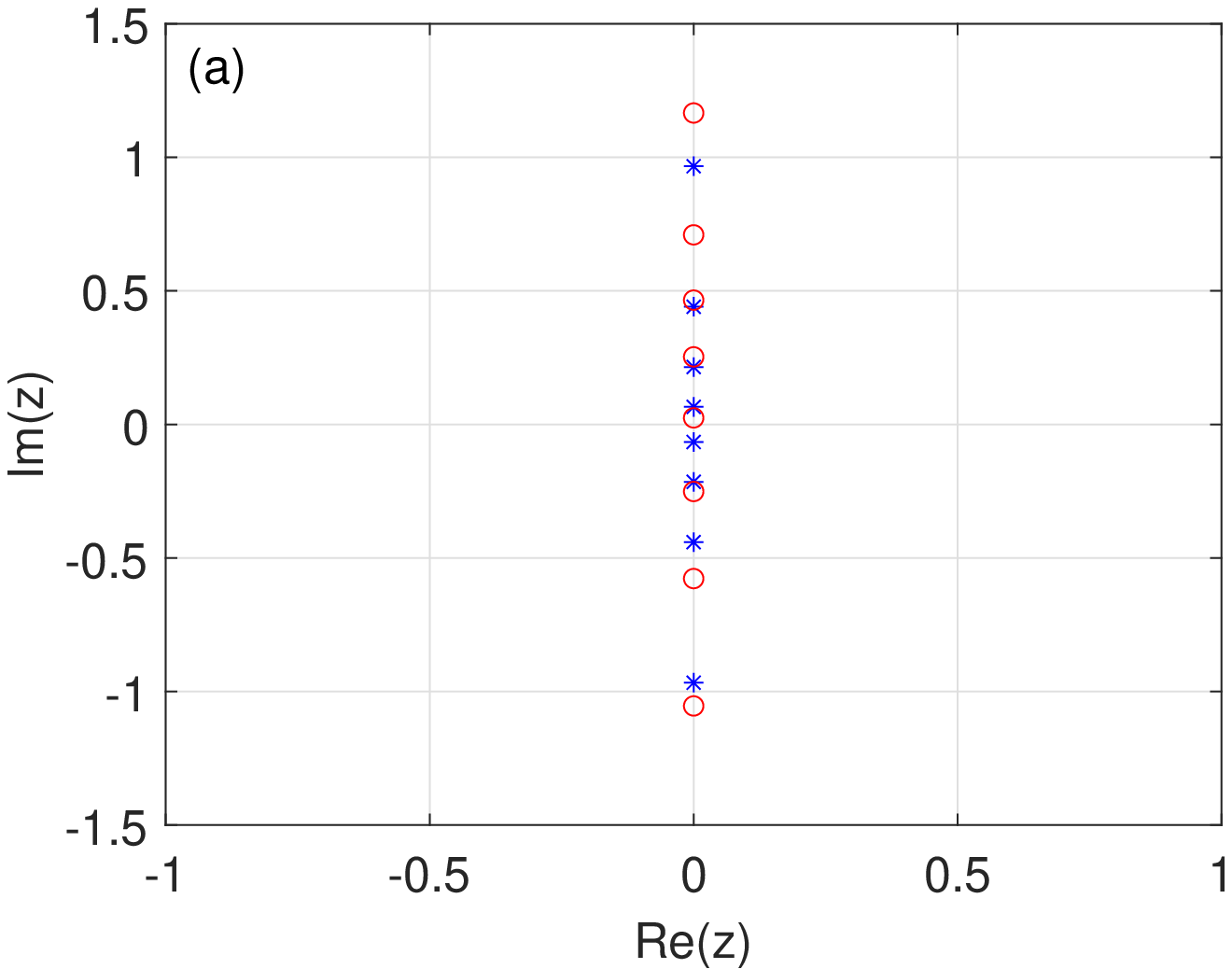}
	\end{minipage}
	\mbox{\hspace{1.50cm}}
	\begin{minipage}[b]{0.40\textwidth}
		\centering
		\includegraphics[height=6cm,width=8cm]{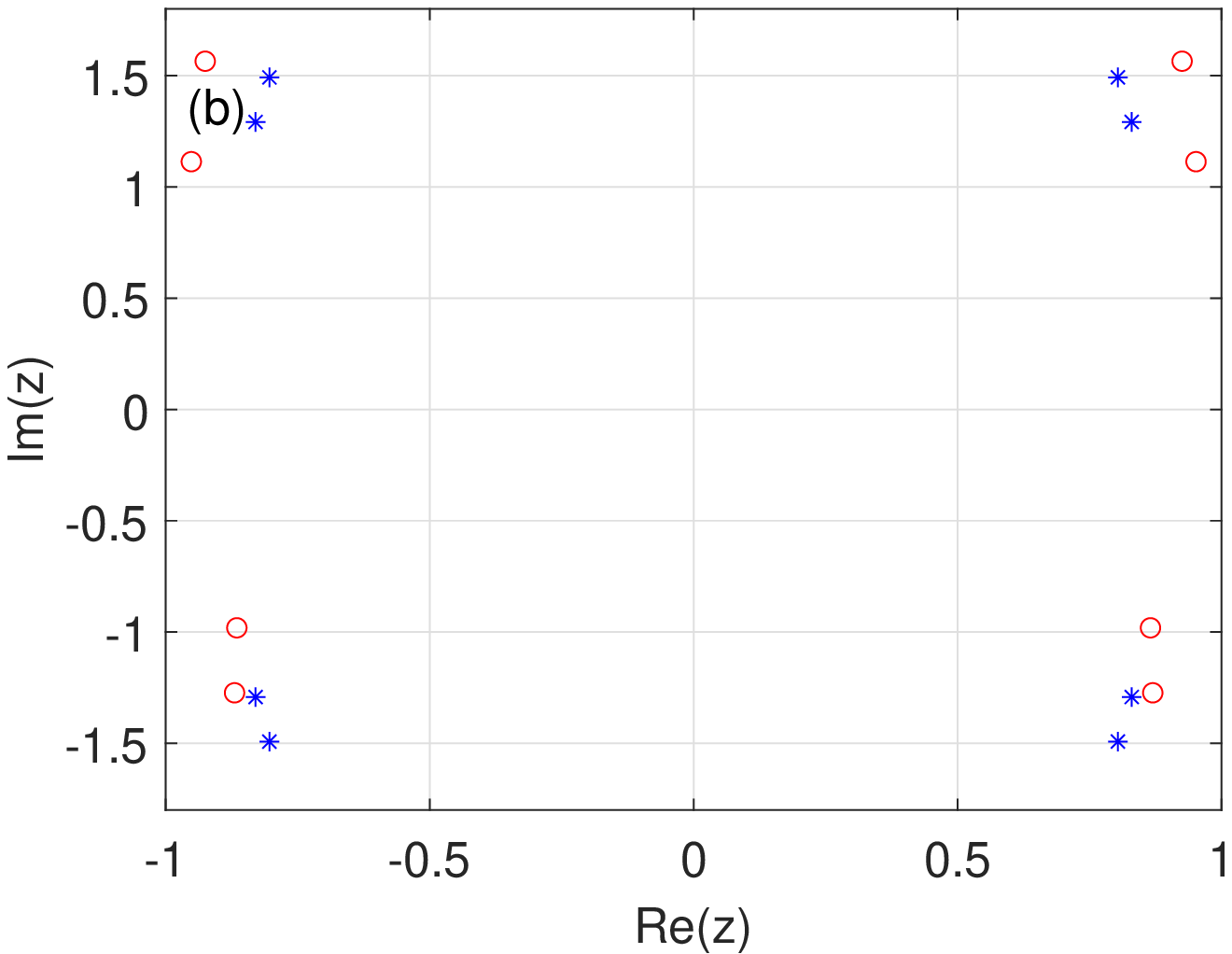}
	\end{minipage}
	\caption{The distribution of zero roots $\{z_j\}$ of $\Lambda(u)$ at the ground state for $N=9$ via exact diagonalization. (a) $\eta=0.75$ and (b) $\eta=0.75+i\pi$.
The blue stars are the results for $\{\theta_{j}=0\}$, and the red circles are the results for arbitrarily chosen inhomogeneity parameters as
$0.14i, 0.32i, -0.43i, 0.54i, -0.25i, 0.63i, 0.47i, -0.78i, 0.19i$. }\label{N9ThetaEta075and075ipiGroundZ-image}
\end{figure*}

Without losing generality, we put $\eta > 0$. Based on the root patterns derived in the previous section, the physical properties such as the ground state energy and the elementary excitations can be computed by adopting the method proposed in \cite{Qiao21} in the thermodynamic limit $N\to\infty$. The key point of this method is to introduce a proper set of inhomogeneity parameters. These parameters serve as an auxiliary tool for analysis and finally will be taken to zero by analytic continuation. For the present case, we choose all the inhomogeneity parameters $\{\theta_j\}$ to be imaginary. Such a choice does not change the patterns of the roots but the root density. Similar analysis for the root patterns with non-zero inhomogeneity parameters can be done by following the same procedure introduced in the previous section. A numerical proof is shown in Fig.\ref{N9ThetaEta075and075ipiGroundZ-image}.

Substituting the ansatz (\ref{lam}) into the functional relations (\ref{TT_relation}), we obtain
\begin{eqnarray}
&&\Lambda^{2}_0\prod_{l=1}^{N-1}\sinh(\theta_j-z_l+\frac{\eta}{2})\sinh(\theta_j-z_l-\frac{\eta}{2})\nonumber \\
&&=-\sinh^{-2N}\eta\prod_{l=1}^{N}\sinh(\theta_j-\theta_l+\eta)\sinh(\theta_j-\theta_l-\eta).\label{TT_relation_concrete}
\end{eqnarray}
Taking the logarithm of the absolute value of Eq.(\ref{TT_relation_concrete}), we have
\begin{eqnarray}\label{TT_relation_abslog}
&&\ln\left|\Lambda^{2}_0\right|+\sum_{l=1}^{N-1}\left[\ln\left|\sinh(\theta_j-z_l+\frac{\eta}{2})\right|+\ln\left|\sinh(\theta_j-z_l-\frac{\eta}{2})\right|\right]\nonumber \\
&&=\ln\left|\sinh^{-2N}\eta\right|+\sum_{l=1}^{N}\left[\ln\left|\sinh(\theta_j-\theta_l+\eta)\right|+\ln\left|\sinh(\theta_j-\theta_l-\eta)\right|\right].
\end{eqnarray}

\subsection{Ground state}

At the ground state, all the $\{\theta_j\}$ and $\{z_l\}$ distribute along the imaginary axis. For convenience, let us put
$\theta_j=i\theta^{\prime}_j$, $ z_l=iz^{\prime}_l$ and $\eta=i\gamma$, where $\theta^{\prime}_j$ and $z^{\prime}_l$ are real and $\gamma$ is imaginary.
In the thermodynamic limit $N\to\infty$, Eq.(\ref{TT_relation_abslog}) becomes
\begin{eqnarray}\label{TT_relation_abslog_tran_limit}
&&\ln\left|\Lambda^{2}_0\right|+N\int_{-\frac{\pi}{2}}^{\frac{\pi}{2}} \ln\left|\sin(\theta-z+\frac{\gamma}{2})\sin(\theta-z-\frac{\gamma}{2})\right|\rho_{1g}(z)dz\nonumber \\
&&=\ln\left|\sinh^{-2N}\eta\right|+N\int_{-\frac{\pi}{2}}^{\frac{\pi}{2}} \ln\left|\sin(\theta-z+\gamma)\sin(\theta-z-\gamma)\right|\sigma(z)dz,
\end{eqnarray}
where $\rho_{1g}(z)$ is the density distribution of $\{z^{\prime}_l\}$, $\sigma(z)$ is the density distribution of $\{\theta^{\prime}_j\}$, replacing $\theta^{\prime}_j$ with $\theta$.
We note that the homogeneous limit $\{\theta_{j}=0\}$ corresponds to $\sigma(z)=\delta(z)$.
Taking the derivative of Eq.(\ref{TT_relation_abslog_tran_limit}), we have
\begin{eqnarray}\label{int_equation}
		\int_{-\frac{\pi}{2}}^{\frac{\pi}{2}}b_{1}(\theta-z)\rho_{1g}(z)dz=\int_{-\frac{\pi}{2}}^{\frac{\pi}{2}}b_{2}(\theta-z)\sigma(z)dz,
\end{eqnarray}
where we define some functions as
\begin{eqnarray}
&&a_{n}(x)=\cot(x-\frac{n\eta i}{2})-\cot(x+\frac{n\eta i}{2}),\label{an}\\
&&b_{n}(x)=\cot(x+\frac{n\eta i}{2})+\cot(x-\frac{n\eta i}{2}),\label{bn}\\
&&c_{n}(x)=\tan(x+\frac{n\eta i}{2})+\tan(x-\frac{n\eta i}{2}).\label{cn}
\end{eqnarray}
In order to solve the integrable equation (\ref{int_equation}), we introduce the Fourier transform
\begin{eqnarray}\label{Fourier_transform}
&&\tilde f(k)=\int_{-\frac{\pi}{2}}^{\frac{\pi}{2}}f(x)e^{-i2kx}dx, \quad k=-\infty, \cdots,+\infty, \nonumber \\
&&f(x)=\frac{1}{\pi}\sum_{k=-\infty}^{+\infty}\tilde f(k)e^{i2kx}, \quad x \in \left[-\frac{\pi}{2},\frac{\pi}{2}\right).
\end{eqnarray}
With the help of the Fourier transform, the integrable equation (\ref{int_equation}) becomes
\begin{eqnarray}\label{int_equation_FT}
	\tilde b_{1}(k)\tilde \rho_{1g}(k)=\tilde b_{2}(k)\tilde\sigma(k),
\end{eqnarray}
where
\begin{eqnarray}
&&\tilde a_{n}(k)=-sign(k)2\pi i e^{-\eta\left|nk\right|},\label{an_ft}\\
&&\tilde b_{n}(k)=2\pi i e^{-\eta\left|nk\right|},\label{bn_ft}\\
&&\tilde c_{n}(k)=(-1)^{k}sign(k)2\pi i e^{-\eta\left|nk\right|}.\label{cn_ft}
\end{eqnarray}
The eigenvalue $\Lambda(u)$ has $N-1$ zero roots, thus the density $\rho_{1g}(z)$ satisfies the normalization $\int_{-\frac{\pi}{2}}^{\frac{\pi}{2}}\rho_{1g}(z)dz=\frac{N-1}{N}$. By taking the homogeneous limit $\sigma(z)=\delta(z)$, we have
 \begin{eqnarray}\label{density distribution k}
 	\tilde\rho_{1g}(k)=\left\{
 	\begin{array}{ll}
 		e^{-\eta\left|k\right|}, & \qquad k=\pm1,\pm2, \cdots,\pm\infty,  \\[6pt]
 		1-\frac{1}{N}, &  \qquad k=0,
 	\end{array}
 	\right.
\end{eqnarray}
and
 \begin{eqnarray}\label{density distribution x}
 	\rho_{1g}(x)=\frac{1}{\pi}\sum_{k=1}^{\infty}2\cos(2kx)e^{-k\eta}+\frac{1}{\pi}\Big(1-\frac{1}{N}\Big).
\end{eqnarray}

\begin{figure*}[t]
	\begin{minipage}[b]{0.40\textwidth}
		\centering
		\includegraphics[height=6cm,width=8cm]{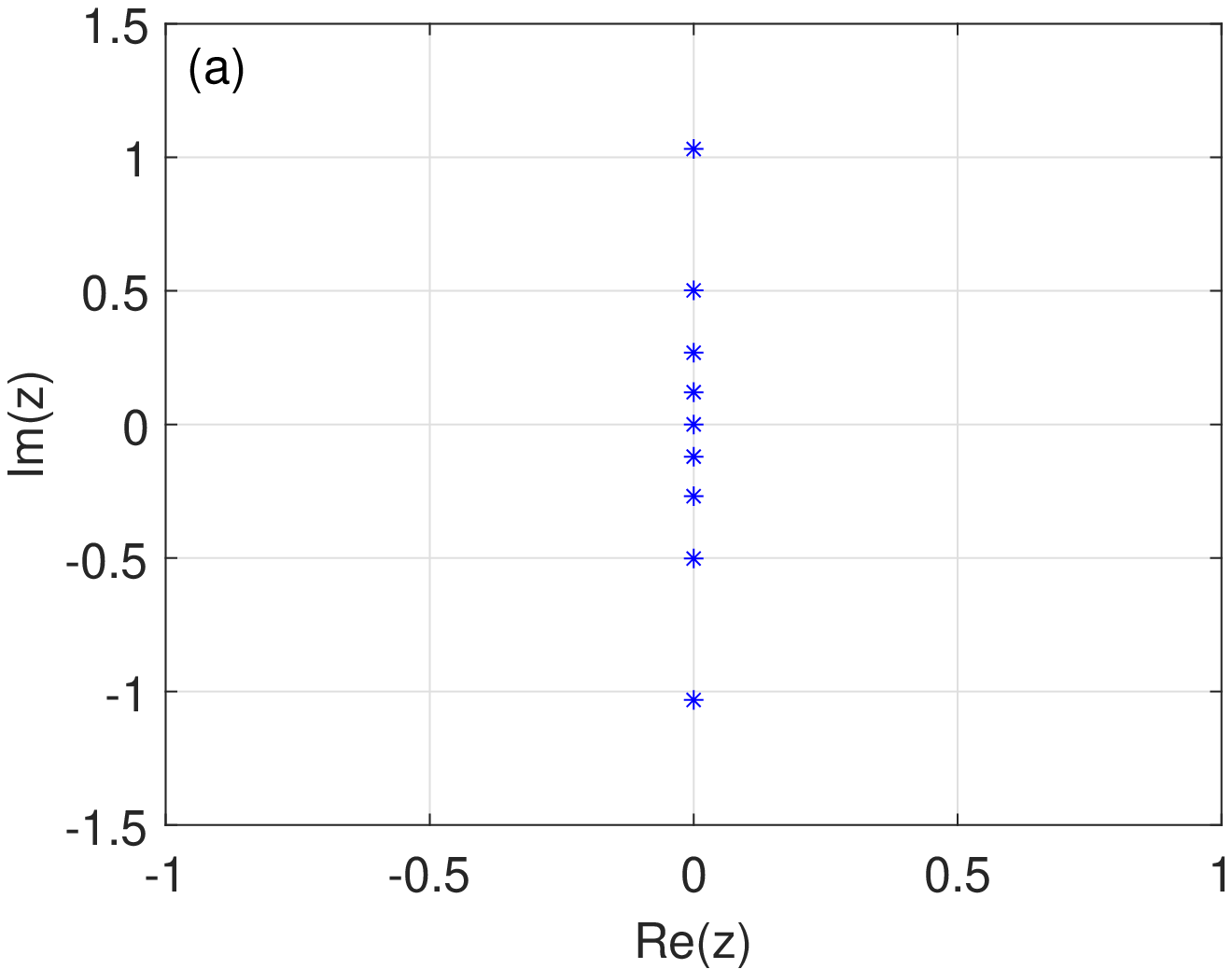}
	\end{minipage}
	\mbox{\hspace{1.50cm}}
	\begin{minipage}[b]{0.40\textwidth}
		\centering
		\includegraphics[height=6cm,width=8cm]{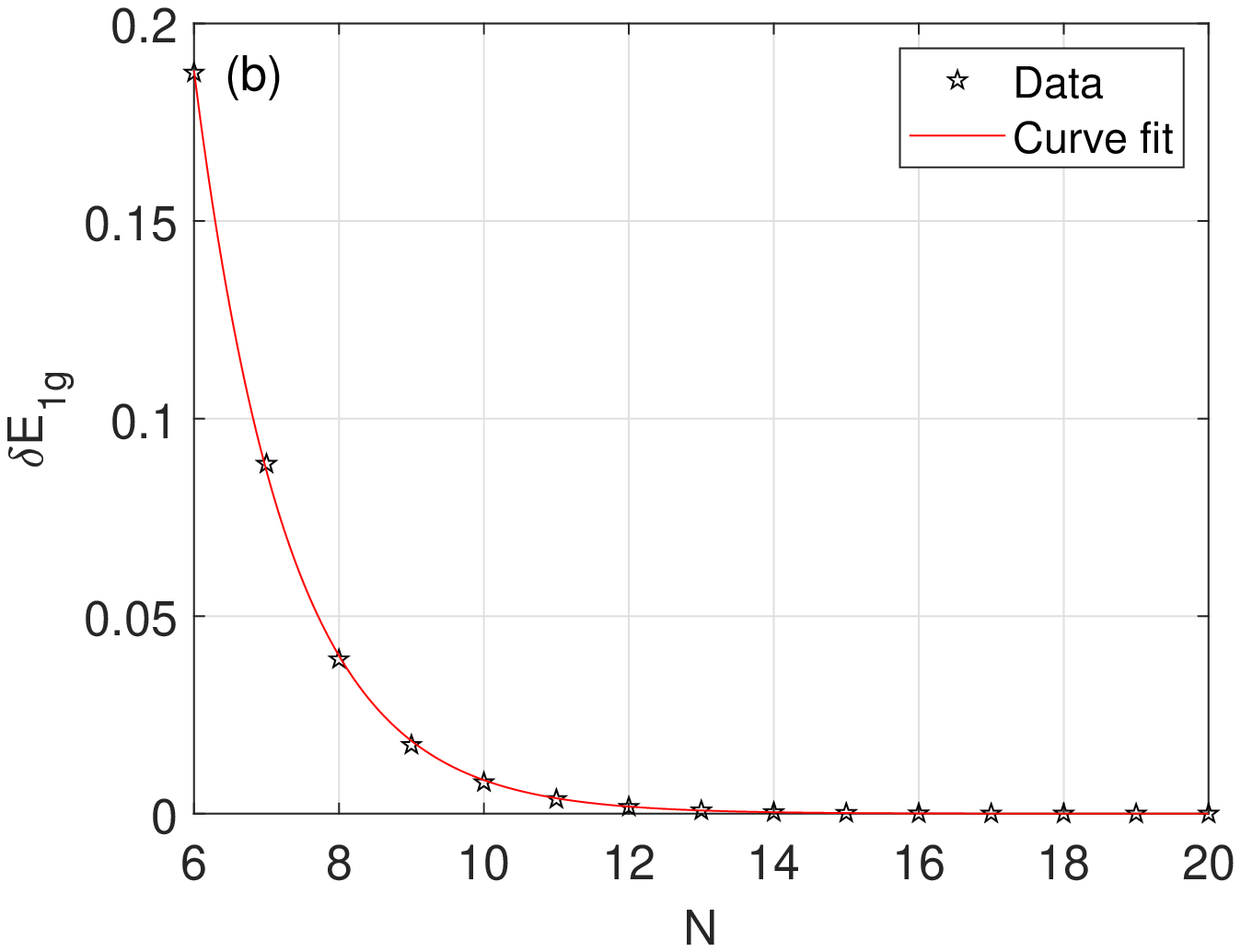}
	\end{minipage}
	\caption{(a) The distribution of zero roots $\{z_j\}$ of $\Lambda(u)$ at the ground state for $N=10$ and $\eta=0.75$ via exact diagonalization.
		(b) The difference $\delta E_{1g}$ between the ground state energy calculated from Eq.(\ref{energy_limit}) and that
		obtained via numerical exact diagonalization of the Hamiltonian (\ref{Ham}). The data can be fitted as $\delta E_{1g}=19.52e^{-0.7738N}$.}\label{EtaRealGroundZDeltaE-image}
\end{figure*}

Thus the ground state energy reads
\begin{eqnarray}\label{energy_limit}
&&	E_{1g}=2N\sinh\eta\int_{-\frac{\pi}{2}}^{\frac{\pi}{2}}\coth(ix-\frac{\eta}{2})\rho_{1g}(x)dx+N\cosh\eta \nonumber \\
&&\quad =-N\cosh\eta+2\sinh\eta.
\end{eqnarray}
This result coincides with the numerical one perfectly as shown in Fig.\ref{EtaRealGroundZDeltaE-image}(b).

\subsection{Elementary excitations}

\begin{figure*}[t]
	\begin{minipage}[b]{0.40\textwidth}
		\centering
		\includegraphics[height=6cm,width=8cm]{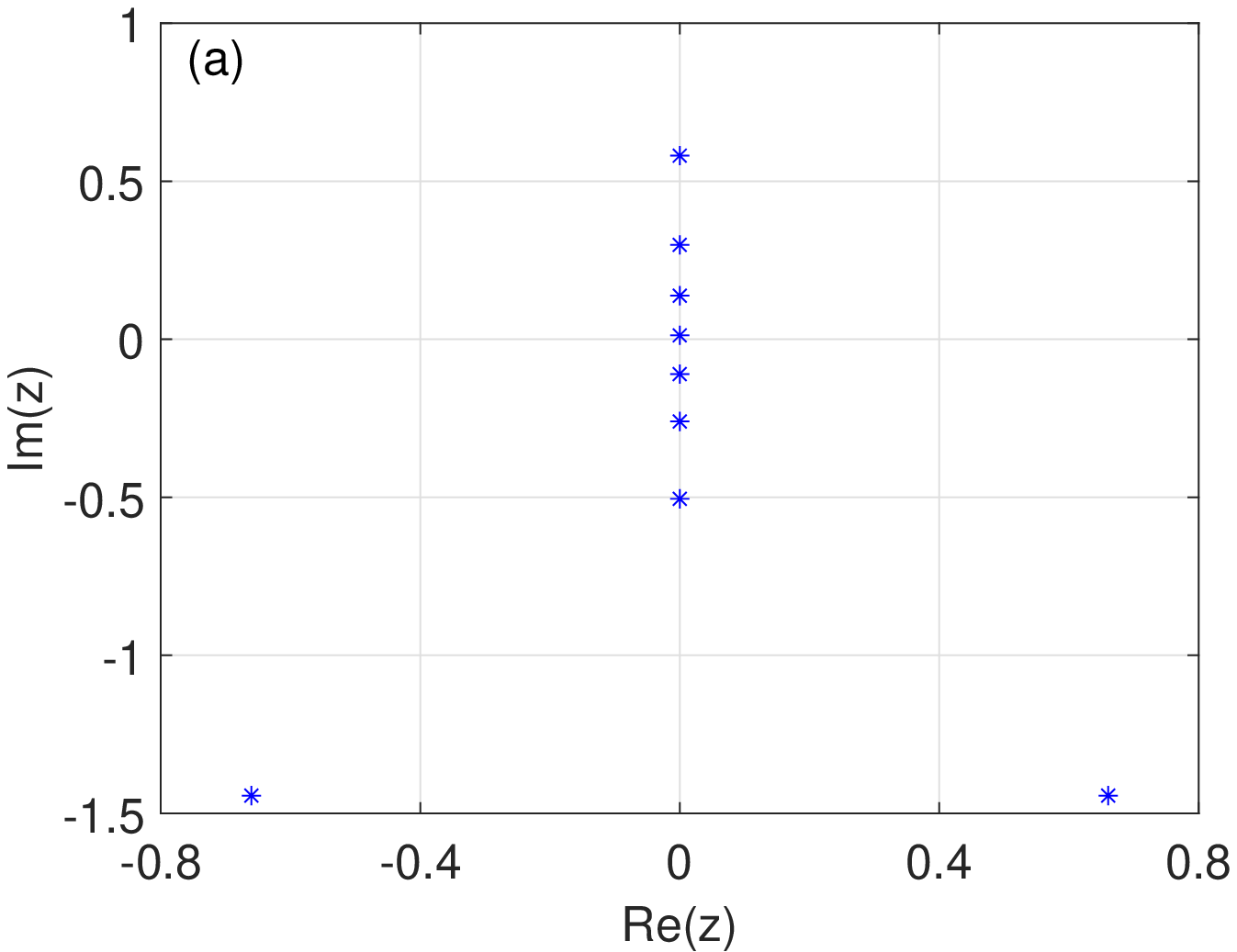}
	\end{minipage}
	\mbox{\hspace{1.50cm}}
	\begin{minipage}[b]{0.40\textwidth}
		\centering
		\includegraphics[height=6cm,width=8cm]{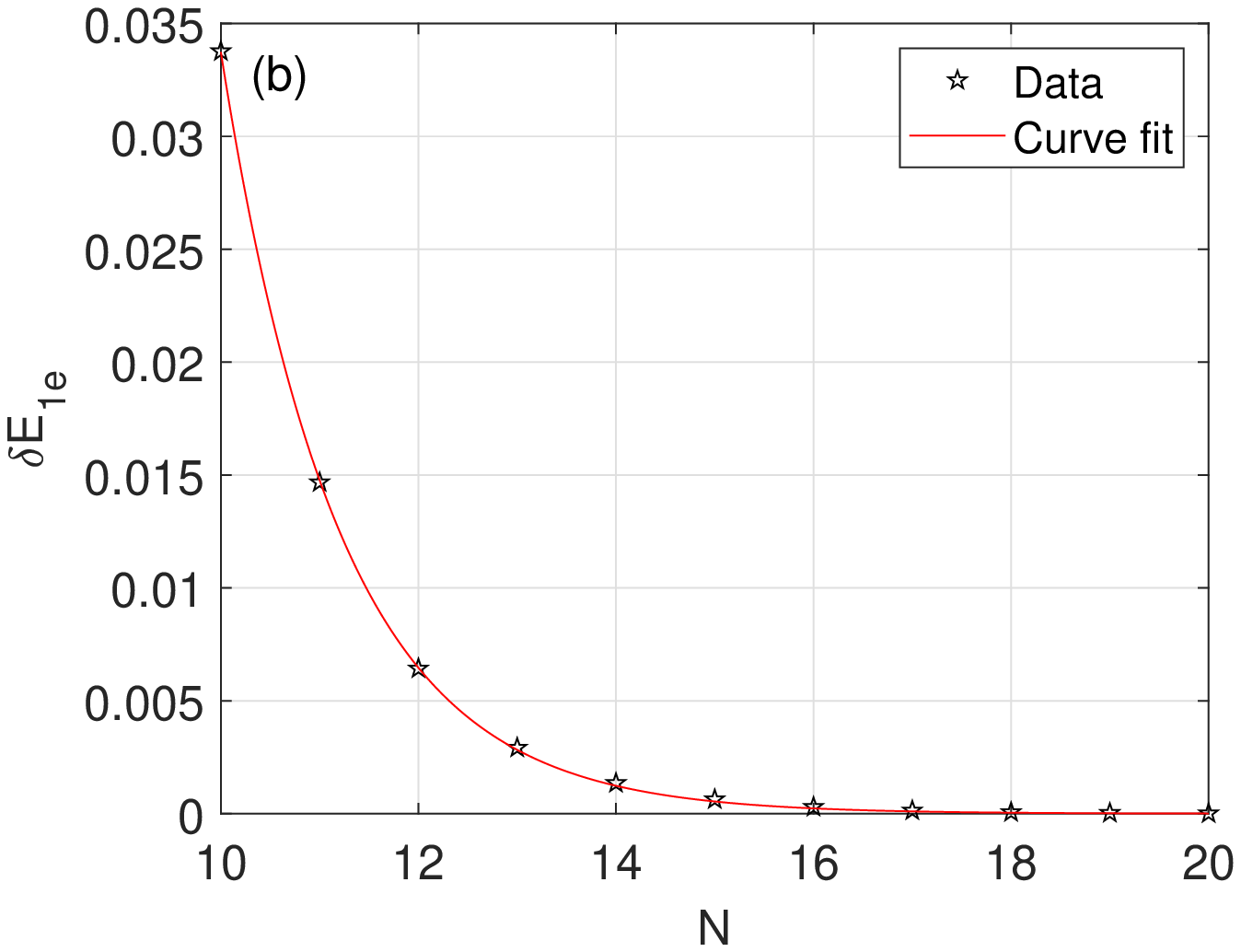}
	\end{minipage}
	\caption{(a) The distribution of zero roots $\{z_j\}$ of $\Lambda(u)$ at the excited state for $N=10$ and $\eta=0.75$. (b)
		The difference $\delta E_{1e}$ between the excited state energy calculated from Eq.(\ref{energy_limit_r_ex}) and that
		obtained by numerical exact diagonalization of the Hamiltonian (\ref{Ham}) with the system-size $N$. The data can be fitted as $\delta E_{1e}=130.4 e^{-0.826 N}$. }\label{EtaRealExcitationZDeltaE-image}
\end{figure*}

Now let us turn to consider the elementary excitations of the system. Due to the root patterns constraints, the excitations can be described by moving several real roots to the complex plane in form of conjugate pairs. The simplest elementary excitation is that two zero roots $z^{\prime}_{N-2}$ and $z^{\prime}_{N-1}$ form a conjugate pair and all the other roots remain real as shown in Fig.\ref{EtaRealExcitationZDeltaE-image}(a).
In this case, the distribution of roots reads
\begin{eqnarray}\label{zdistribution}
	z_l=iz^{\prime}_l,{~~}(l=1,\cdots,N-3),\nonumber\\
z_{N-2}=iz^{\prime}_{N-2}=\frac{n\eta}{2}+i\alpha+o(e^{-\delta N}),\nonumber\\
 z_{N-1}=iz^{\prime}_{N-1}=-\frac{n\eta}{2}+i\alpha+o(e^{-\delta N}),
\end{eqnarray}
where $z^{\prime}_l$ and $\alpha$ are real and $n\ge2$.

Substituting Eq.(\ref{zdistribution}) into (\ref{TT_relation_abslog}), we have
\begin{eqnarray}\label{TT_relation_r_ex}
&&\ln\left|\Lambda^{2}_0\right|+\sum_{l=1}^{N-3}\left[\ln\left|\sinh(i\theta^{\prime}_j-iz^{\prime}_l+\frac{i\gamma}{2})\right|+\ln\left|\sinh(i\theta^{\prime}_j-iz^{\prime}_l-\frac{i\gamma}{2})\right|\right]\nonumber \\
&&\quad +\left[\ln\left|\sinh(i\theta^{\prime}_j-(\frac{n\eta}{2}+i\alpha)+\frac{i\gamma}{2})\right|+\ln\left|\sinh(i\theta^{\prime}_j-(\frac{n\eta}{2}+i\alpha)-\frac{i\gamma}{2})\right|\right]\nonumber \\
&&\quad +\left[\ln\left|\sinh(i\theta^{\prime}_j-(-\frac{n\eta}{2}+i\alpha)+\frac{i\gamma}{2})\right|+\ln\left|\sinh(i\theta^{\prime}_j-(-\frac{n\eta}{2}+i\alpha)-\frac{i\gamma}{2})\right|\right]\nonumber \\
&&=\ln\left|\sinh^{-2N}\eta\right|+\sum_{l=1}^{N}\left[\ln\left|\sinh(i\theta^{\prime}_j-i\theta^{\prime}_l+i\gamma)\right|+\ln\left|\sinh(i\theta^{\prime}_j-i\theta^{\prime}_l-i\gamma)\right|\right].
\end{eqnarray}
In the thermodynamic limit $N\to\infty$, the Fourier transformation of Eq.(\ref{TT_relation_r_ex}) gives
\begin{eqnarray}\label{int_equation_r_ex_FT}
	N\tilde b_{1}(k)\tilde\rho_{1e}(k)+e^{-i2k\alpha}\tilde b_{n-1}(k)+e^{-i2k\alpha}\tilde b_{n+1}(k)=N\tilde b_{2}(k)\tilde\sigma(k).
\end{eqnarray}
The solution of Eq.(\ref{int_equation_r_ex_FT}) reads
 \begin{eqnarray}\label{density distribution k_r_ex}
	\tilde\rho_{1e}(k)=\left\{
	\begin{array}{ll}
		e^{-\eta\left|k\right|}-\frac{e^{-i2k\alpha}}{N}(e^{-n\eta\left|k\right|}+e^{-(n-2)\eta\left|k\right|}), &  k=\pm1,\pm2, \cdots,\pm\infty, \\[6pt]
		1-\frac{3}{N}, &  k=0,
	\end{array}
	\right.
\end{eqnarray}
and
\begin{eqnarray}\label{density distribution x_r_ex}
\rho_{1e}(x)=\frac{1}{\pi}\sum_{k=1}^{\infty}\left[2\cos(2kx)e^{-k\eta}-2\cos[2k(x-\alpha)]\frac{e^{-n\eta k}+e^{-(n-2)\eta k}}{N}\right]+\frac{1}{\pi}\Big(1-\frac{3}{N}\Big).
\end{eqnarray}

The energy at this excited state is characterized by the density of roots $\rho_{1e}(x)$ as
\begin{eqnarray}\label{energy_limit_r_ex}
&&E_{1e}=2N\sinh\eta\int_{-\frac{\pi}{2}}^{\frac{\pi}{2}}\coth(ix-\frac{\eta}{2})\rho_{1e}(x)dx+N\cosh\eta \nonumber \\
&&\qquad +2\sinh\eta\left[\coth(\frac{n\eta}{2}+i\alpha-\frac{\eta}{2})+\coth(-\frac{n\eta}{2}+i\alpha-\frac{\eta}{2})\right].
\end{eqnarray}
The energy carried by this elementary excitation is
\begin{eqnarray}\label{energy_gap_r}
		\Delta E_{1}=4\sinh\eta\frac{\sinh\left[(n-1)\eta\right]}{\cosh\left[(n-1)\eta\right]-2\cos(2\alpha)}.
\end{eqnarray}
If $n=2$ and $\alpha=\pm\frac{\pi}{2}$, the energy arrives at its minimum value,
\begin{eqnarray}\label{energy_gap_min_r}
	\Delta E_{1min}=4\sinh\eta\tanh\frac{\eta}{2}.
\end{eqnarray}
Comparison of our analytic results and numerical results is shown in Fig.\ref{EtaRealExcitationZDeltaE-image}(b). This result Eq.(\ref{energy_gap_min_r}) also coincides with that given in \cite{Qiao18}.

\section{Exact solution for $\eta \in \mathbb{R}+i\pi$}\label{5}
\setcounter{equation}{0}
For convenience, we put $\eta=\eta_++i\pi$ with $\eta_+$ a real number.
\subsection{Even $N$ case}
\begin{figure*}[t]
	\begin{minipage}[b]{0.40\textwidth}
		\centering
		\includegraphics[height=6cm,width=8cm]{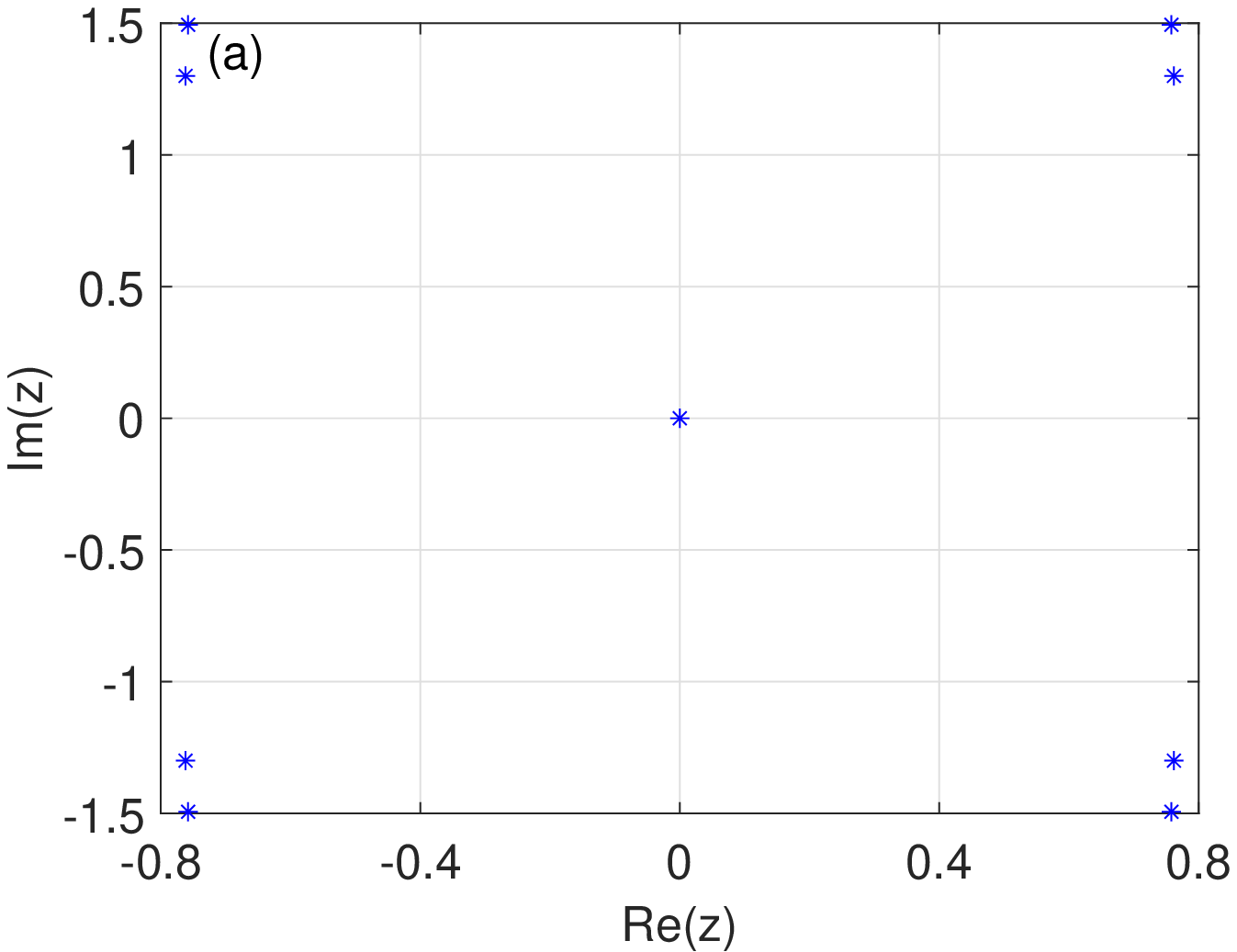}
	\end{minipage}
	\mbox{\hspace{1.50cm}}
	\begin{minipage}[b]{0.40\textwidth}
		\centering
		\includegraphics[height=6cm,width=8cm]{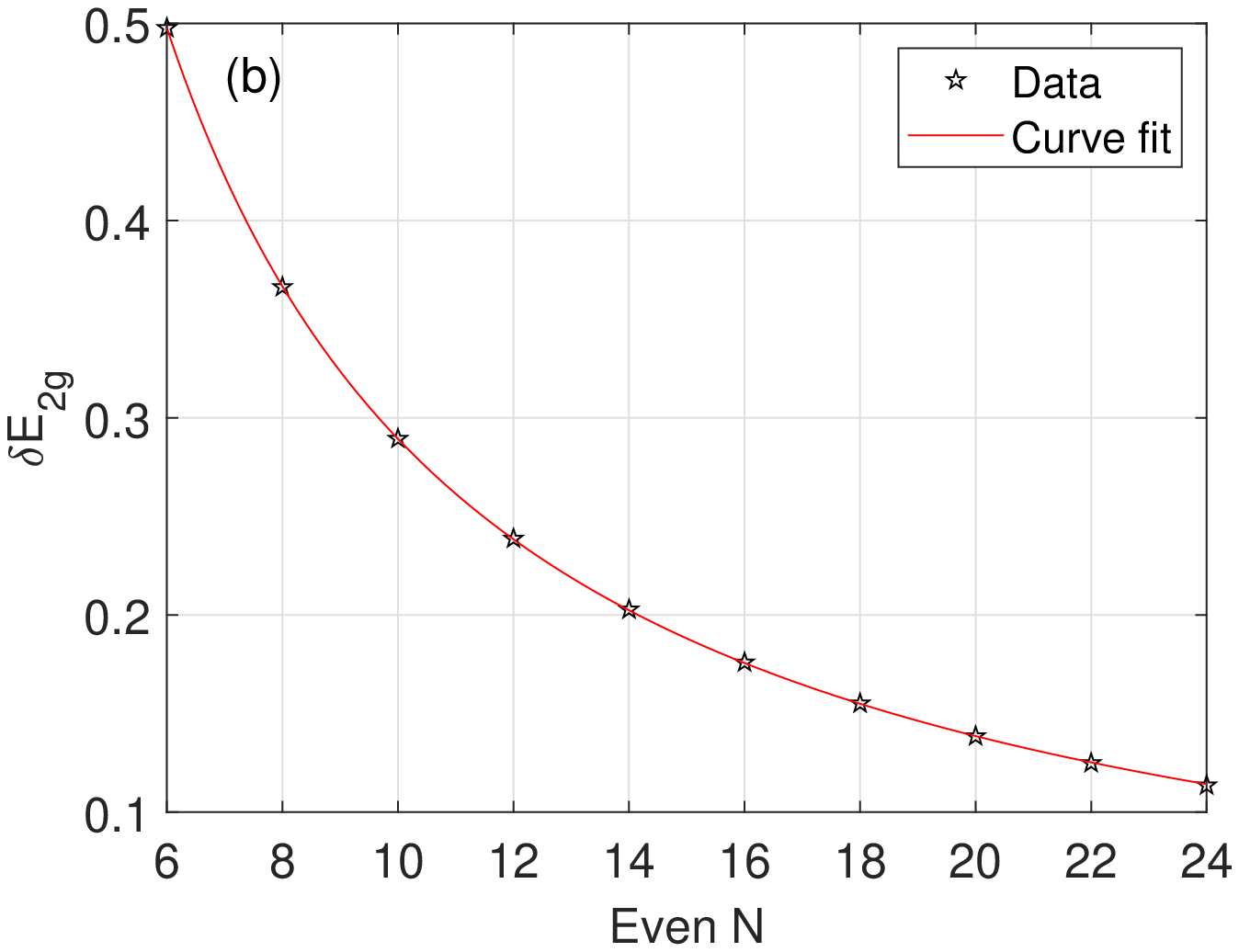}
	\end{minipage}
	\caption{(a) The distribution of $z$ zero roots at the ground state for $N=10$ and $\eta=0.75+i\pi$. (b)
The difference $\delta E_{2g}$ between the ground state energy calculated from Eq.(\ref{energy_ground_im_even}) and that obtained via numerical exact diagonalization. The data can be fitted as $\delta E_{2g}=3.429 N^{-1.073}$.}\label{EtaRealipiNEvenGroundZDeltaE-image}
\end{figure*}

We first consider the case of even $N$. In this case, the number of roots $N-1$ is an odd number. At the ground state, due to the root patterns constraints, the root patterns read
\begin{eqnarray}\label{zdistribution_im}
	z_l=\eta_{+}+iz^{\prime}_l, \quad
	z_{l+\frac{N}{2}-1}=-\eta_{+}+iz^{\prime}_l, \quad l=1,\cdots,\frac{N}{2}-1, \quad z_{N-1}=i\beta,
\end{eqnarray}
where $\eta=\eta_{+}+i\pi$, $\eta_{+}$, $z^{\prime}_l$ and $\beta$ are real. The variation of the real root $\beta$ gives the gapless excitation. At the ground state
$\beta=0$, while at the excited state $\beta\neq0$. A numerical result for $N=10$ is shown in Fig.\ref{EtaRealipiNEvenGroundZDeltaE-image}(a).

\begin{figure}[t]
	\centering
	\includegraphics[height=9cm,width=12cm]{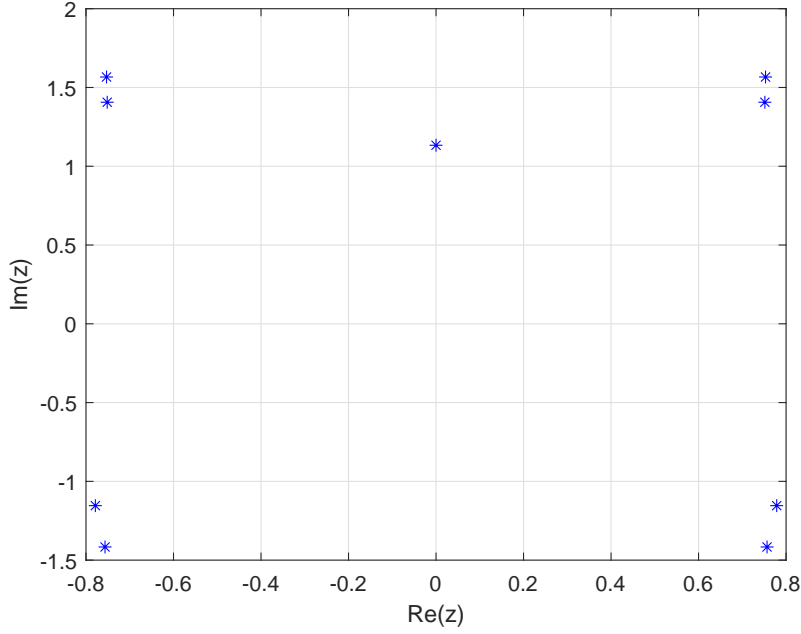}
	\caption{ A distribution of zero roots $\{z_l\}$ for the ground state and the low-lying excited state for $N=10$ and $\eta=0.75+i\pi$.}\label{EtaRealipiNEvenExcitationZ-image}
\end{figure}

With the same procedure as before, substituting the patterns of zero roots into Eq.(\ref{TT_relation_abslog}) and considering the thermodynamic limit, we
obtain that the densities of $z^{\prime}_l$ satisfy
\begin{eqnarray}\label{int_equation_im_even}
	-N\left[\tilde c_{1}(k)+\tilde c_{3}(k)\right]\tilde\rho_{2e}(k)-e^{-i2k\beta}\tilde c_{1}(k)=N\tilde b_{2}(k)\tilde\sigma(k),
\end{eqnarray}
where $\tilde\rho_{2e}(k)$ and $\tilde\sigma(k)$ are the Fourier transformations of the density of zero roots $z^{\prime}_l$ and that of the inhomogeneity parameters, respectively.
We note that the $\rho_{2e}(z)$ is the density of $z^{\prime}_l$ and satisfies the normalization $\int_{-\frac{\pi}{2}}^{\frac{\pi}{2}}\rho_{2e}(z)dz=\frac{1}{2}-\frac1{N}$.
The solution of Eq.(\ref{int_equation_im_even}) is
\begin{eqnarray}\label{density distribution k_im_even}
	\tilde\rho_{2e}(k)=\left\{
	\begin{array}{ll}
	-\frac{\frac{1}{N}e^{-i2k\beta}-(-1)^{k}e^{-\eta_{+}k}}{1+e^{-2\eta_{+}k}}, & k=1,2, \cdots,\infty,  \\[8pt]
	\frac12-\frac{1}{N}, & k=0, \\[8pt]
	-\frac{\frac{1}{N}e^{-i2k\beta}+(-1)^{k+1}e^{\eta_{+}k}}{1+e^{2\eta_{+}k}}, & k=-1,-2, \cdots,-\infty,
	\end{array}
	\right.
\end{eqnarray}
and
\begin{eqnarray}\label{density distribution x_im_even} \rho_{2ge}(x)=\frac{1}{\pi}\sum_{k=1}^{\infty}\left[2\cos(2kx)\frac{(-1)^{k}e^{-\eta_{+}k}}{1+e^{-2\eta_{+}k}}-2\cos[2k(x-\beta)]\frac{\frac{1}{N}}{1+e^{-2\eta_{+}k}}\right]+\frac{1}{\pi}\Big(\frac{N-2}{2N}\Big).
\end{eqnarray}
The eigenenergy can be calculated as
\begin{eqnarray}\label{energy_limit_im_even}
&& E_{2e}=2N\sinh\eta\int_{-\frac{\pi}{2}}^{\frac{\pi}{2}}\left[\coth(\eta_{+}+ix-\frac{\eta}{2})+\coth(-\eta_{+}+ix-\frac{\eta}{2})\right]\rho_{2e}(x)dx \nonumber \\
&&\qquad +2\sinh\eta\coth(i\beta-\frac{\eta}{2})+N\cosh\eta.
\end{eqnarray}
For the ground state, $\beta=0$ and corresponding energy is
\begin{eqnarray}\label{energy_ground_im_even}
&&E_{2g}=-4N\sinh\eta_{+}\sum_{k=1}^{\infty}e^{-2\eta_{+}k}\tanh(\eta_{+}k)-4\sinh\eta_{+}\sum_{k=1}^{\infty}(-1)^{k+1}e^{-\eta_{+}k}\tanh(\eta_{+}k)\nonumber \\
&&\qquad\;\; +2\sinh\eta_{+}\tanh\Big(\frac{\eta_{+}}{2}\Big)-N\cosh\eta_{+}.
\end{eqnarray}
For the simplest excited state, $\beta\neq0$ as shown in Fig.\ref{EtaRealipiNEvenExcitationZ-image} for finite $N$.
After tedious calculation, we find the energy difference $\Delta E_{2}=E_{2e}-E_{2g}$ as
\begin{eqnarray}\label{energy_ex_im_even}
&&\Delta E_{2}=-4\sinh\eta_{+}\sum_{k=1}^{\infty}(-1)^{k+1}e^{-\eta_{+}k}\tanh(\eta_{+}k)[\cos(2k\beta)-1]\nonumber \\
&&\qquad\quad -2\sinh\eta_{+}\Big(\tanh\frac{\eta_{+}}{2}-\frac{\sinh\eta_{+}}{\cosh\eta_{+}+\cos2\beta}\Big).
\end{eqnarray}
Detailed analysis of Eq.(\ref{energy_ex_im_even}) shows that $\Delta E \rightarrow 0$ when $\beta\rightarrow0$, which means that the elementary excitation is gapless for an even $N$.

\subsection{Odd $N$ case}
\begin{figure*}[t]
	\begin{minipage}[b]{0.40\textwidth}
		\centering
		\includegraphics[height=6cm,width=8cm]{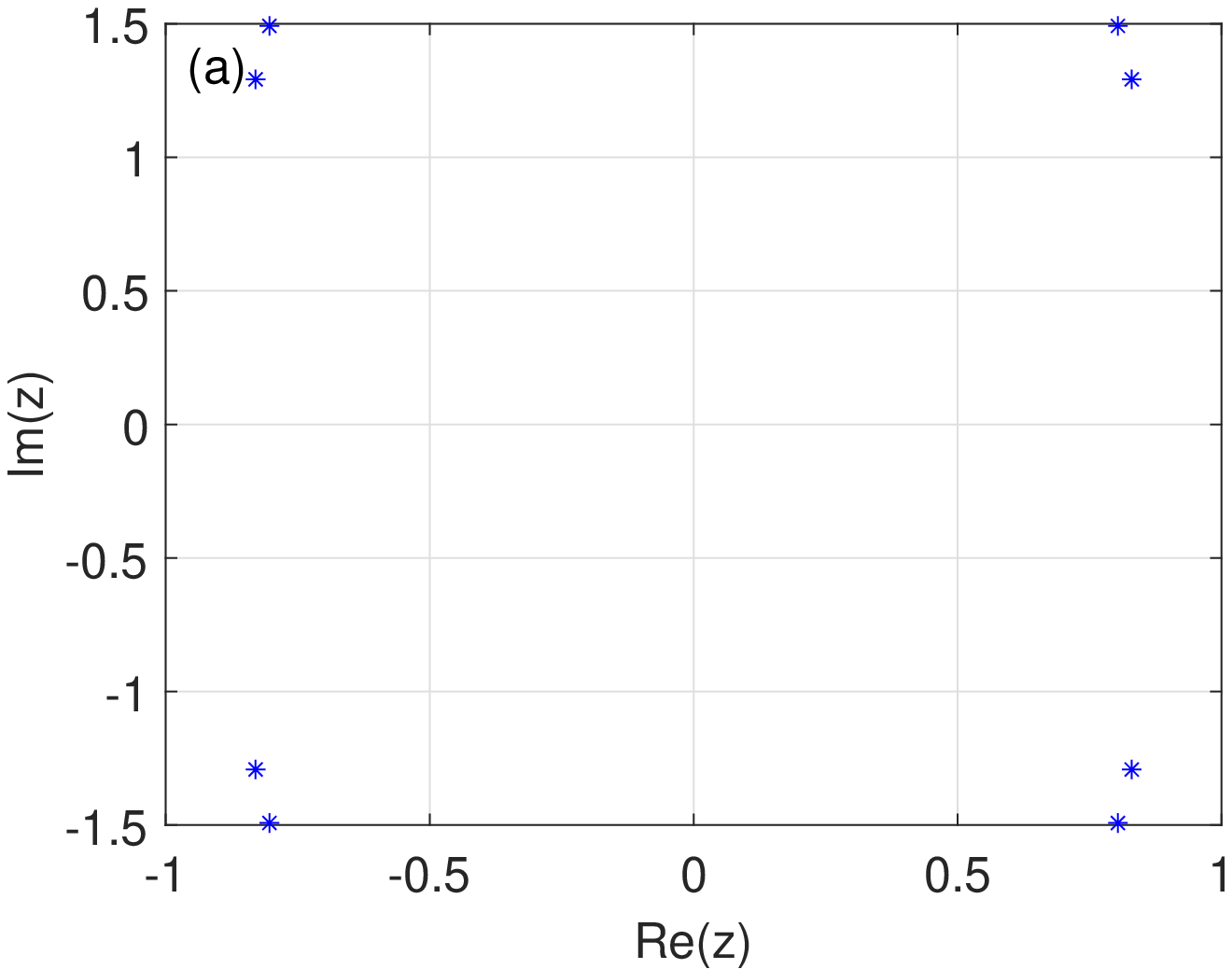}
	\end{minipage}
	\mbox{\hspace{1.50cm}}
	\begin{minipage}[b]{0.40\textwidth}
		\centering
		\includegraphics[height=6cm,width=8cm]{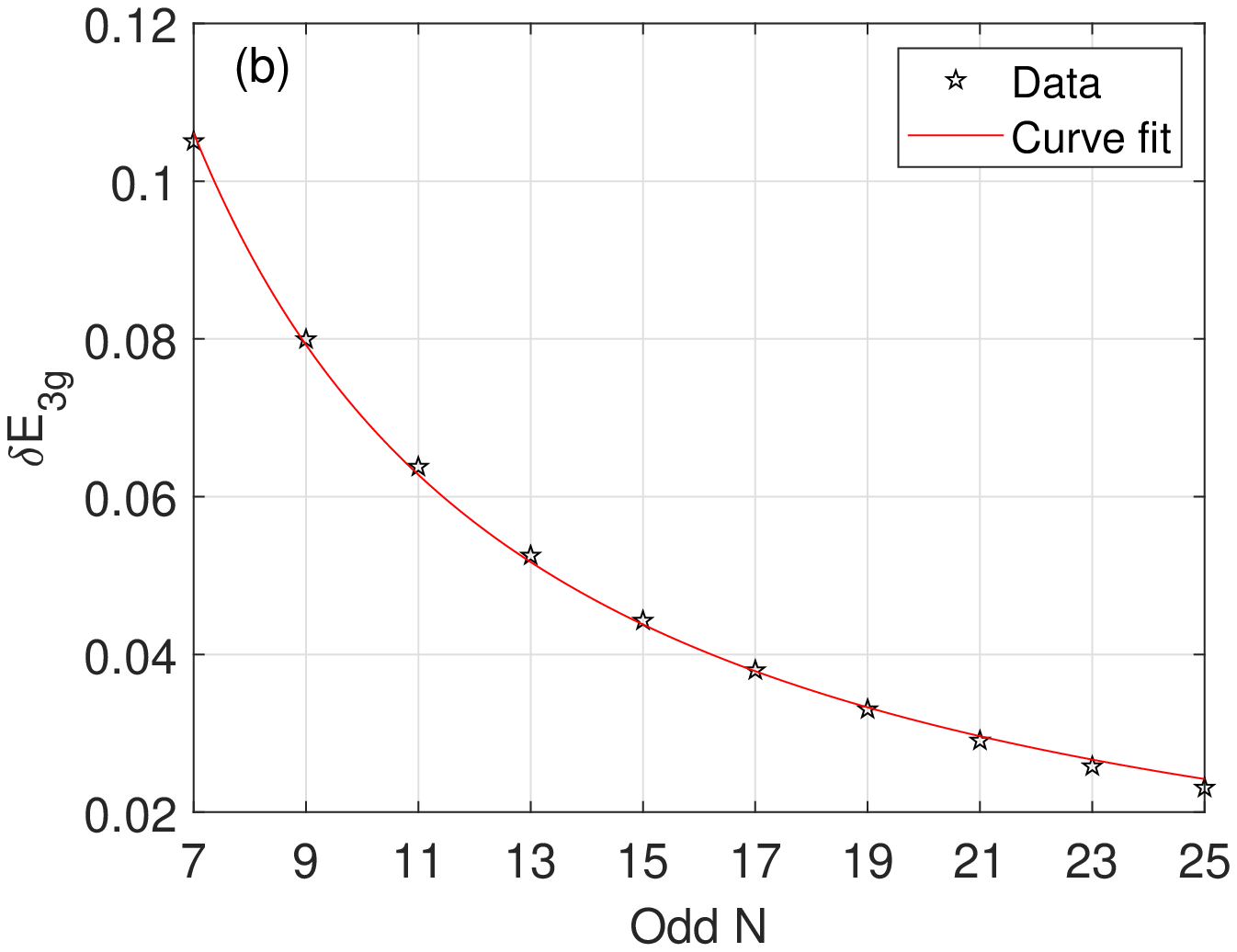}
	\end{minipage}
	\caption{(a) The distribution of $z$ roots at the ground state for $N=9$ and $\eta=0.75+i\pi$. (b) The difference $\delta E_{3g}$ between the ground state energy calculated from Eq.(\ref{energy_ground_im_odd}) and that obtained via numerical exact diagonalization. The data can be fitted as $\delta E_{3g}=1.016 N^{-1.161}$.}\label{EtaRealipiNOddGroundZDeltaE-image}
\end{figure*}

For an odd $N$, the number of roots is even and all the $\{z_j'\}$ roots form conjugate pairs in the ground state as shown in Fig.\ref{EtaRealipiNOddGroundZDeltaE-image}(a). The distribution of zero roots for the ground state is
 \begin{eqnarray}\label{zdistribution_im_odd}
 	z_l=\eta_{+}+iz^{\prime}_l, \quad
 	z_{l+\frac{N-1}{2}}=-\eta_{+}+iz^{\prime}_l, \quad l=1,\cdots,\frac{N-1}{2}.
 \end{eqnarray}
Repeating the previous procedure we obtain the density of roots in the momentum space as
 \begin{eqnarray}\label{density distribution k_im_odd}
 	\tilde\rho_{3g}(k)=\left\{
 	\begin{array}{ll}
 		\frac{(-1)^{k}e^{-\eta_{+}k}}{1+e^{-2\eta_{+}k}}, & k=1,2, \cdots,\infty , \\[8pt]
 		\frac{N-1}{2N}, & k=0, \\[8pt]
 		\frac{(-1)^{k}e^{\eta_{+}k}}{1+e^{2\eta_{+}k}}, & k=-1,-2, \cdots,-\infty.
 	\end{array}
 	\right.
 \end{eqnarray}
With the help of Fourier transformation, the density of $z^{\prime}_l$ can be obtained as
 \begin{eqnarray}\label{density distribution x_im_odd}
 	\rho_{3g}(x)=\frac{1}{\pi}\sum_{k=1}^{\infty}\left[2\cos2kx\frac{(-1)^{k}e^{-\eta_{+}k}}{1+e^{-2\eta_{+}k}}\right]+\frac{1}{\pi}\Big(\frac{N-1}{2N}\Big).
 \end{eqnarray}
The ground state energy is
\begin{eqnarray}\label{energy_ground_im_odd}
 		E_{_{3g}}=-4N\sinh\eta_{+}\sum_{k=1}^{\infty}e^{-2\eta_{+}k}\tanh(\eta_{+}k)-N\cosh\eta_{+}.
\end{eqnarray}

\begin{figure}[t]
	\centering
	\includegraphics[height=9cm,width=12cm]{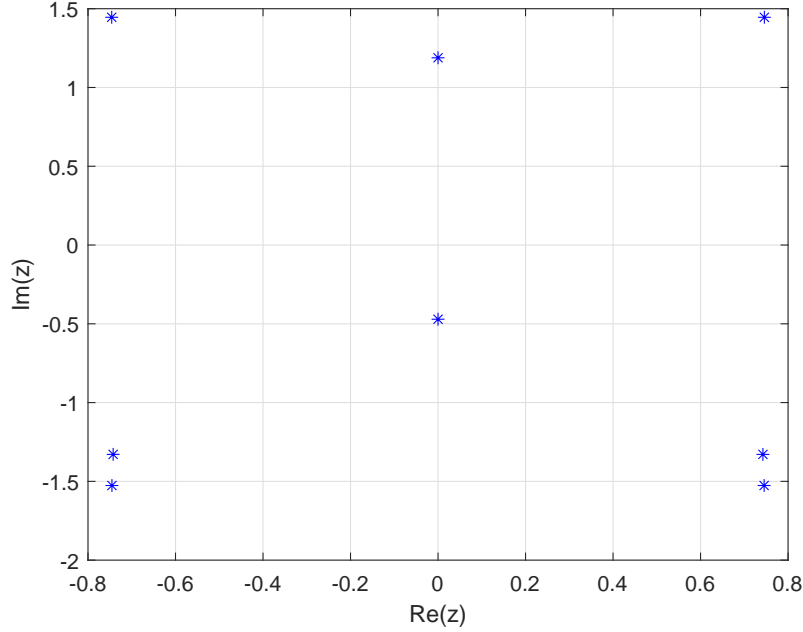}
	\caption{ The distribution of roots $\{z_l\}$  at a low-lying excited state for $N=9$ and $\eta=0.75+i\pi$.}\label{EtaRealipiNOddExcitationZ-image}
\end{figure}

A low-lying excited state can be describe by root patterns
\begin{eqnarray}\label{zdistribution_im_odd_ex}
&& z_l=\eta_{+}+iz^{\prime}_l, \quad z_{l+\frac{N-3}{2}}=-\eta_{+}+iz^{\prime}_l, \quad l=1,\cdots,\frac{N-3}{2},\nonumber \\
&&    z_{N-2}=ip, \quad z_{N-1}=iq,
 \end{eqnarray}
as shown in Fig.\ref{EtaRealipiNOddExcitationZ-image} for $N=9$, where $p$ and $q$ are real. In the thermodynamic limit, the density of zero roots reads
 \begin{eqnarray}\label{density distribution x_im_odd_ex}
&&\rho_{3e}(x)=\frac{1}{\pi}\sum_{k=1}^{\infty} \left\{-2\left\{\cos[2k(x-p)]+\cos[2k(x-q)]\right\}\frac{1}{N(1+e^{-2\eta_{+}k})} \right.\nonumber \\
&&\qquad\quad \left.+2\cos(2kx)\frac{(-1)^{k}e^{-\eta_{+}k}}{1+e^{-2\eta_{+}k}}\right\}+\frac{1}{\pi}\Big(\frac{N-3}{2N}\Big).
  \end{eqnarray}
The excitation energy is given by
\begin{eqnarray}\label{energy_ex_im_ex_gap}
		\Delta E_{3}=\epsilon(p)+\epsilon(q),
\end{eqnarray}
with
\begin{eqnarray}\label{energy_ex_im_ex_gap_e0} \epsilon(t)=-4\sinh\eta_{+}\sum_{k=1}^{\infty}(-1)^{k+1}e^{-\eta_{+}k}\tanh(\eta_{+}k)\cos(2kt)+2\sinh\eta_{+}\frac{\sinh\eta_{+}}{\cosh\eta_{+}+\cos2t}.
\end{eqnarray}
The excitation energy reaches its minimum at the point of $p=q=0$ and the value is
\begin{eqnarray}\label{energy_ex_im_ex_gap_min}
\Delta E_{3min}=-8\sinh\eta_{+}\sum_{k=1}^{\infty}(-1)^{k+1}e^{-\eta_{+}k}\tanh(\eta_{+}k)+4\sinh\eta_{+}\tanh\frac{\eta_{+}}{2}.
\end{eqnarray}

Interestingly, our result for odd $N$ coincides perfectly with those calculated via density matrix renormalization group method \cite{White93,Schollwck05} for even $N$ periodic chain.

\begin{figure}[t]
	\centering
	\includegraphics[height=9cm,width=12cm]{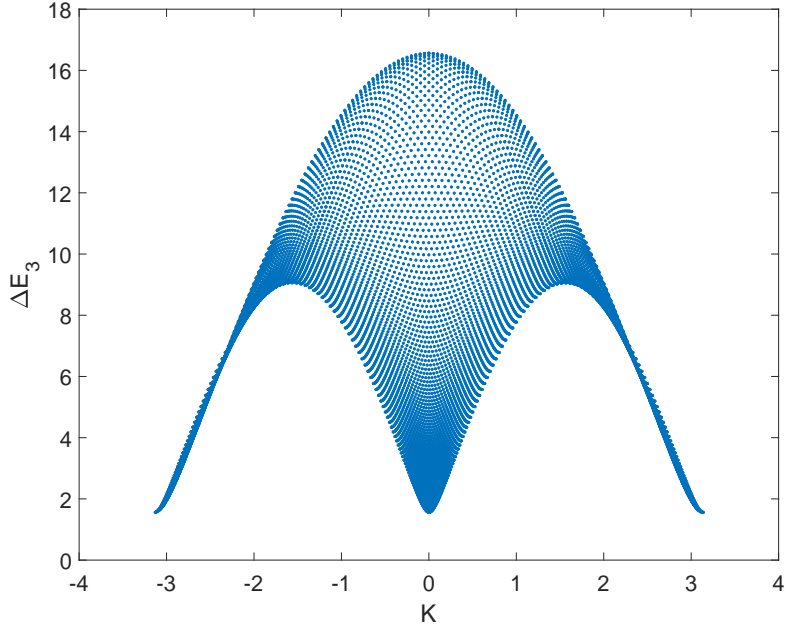}
	\caption{ The dispersion relation of a single excitation for $\eta=1.31696+i\pi$. }\label{Dispersion relation 3-image}
\end{figure}

Despite the absence of translational invariance in the present model, a topological momentum operator can be defined \cite{Qiao20}. We note that the $t(0)$ is a conserved quantity and $t^{2N}(0)=1$ \cite{Book}. Therefore, $t(0)$ can be treated as a shift operator in the $Z_2$ topological manifold, which allow us to define the topological momentum operator as $\hat k=-i\ln t(0)$ with the eigenvalue
 \begin{eqnarray}\label{momentum_def}
		k=-i\ln\Lambda(0).
\end{eqnarray}
Substituting the value of $\Lambda(0)$ into Eq.(\ref{momentum_def}), we obtain the momentum
 \begin{eqnarray}\label{momentum_exp}
		k=-\frac{i}{2}\ln\prod_{l=1}^{N-1}\frac{\sinh(z_{l}+\frac{\eta}{2})}{\sinh(z_{l}-\frac{\eta}{2})}+\frac{\pi}{4}(1-(-1)^{N-1}),
\end{eqnarray}
which is also determined by the zero roots $\{z_l\}$. In the thermodynamic limit, the momentum reads
 \begin{eqnarray}\label{momentum_con}
 &&k=-\frac{i}{2}N\int_{-\frac{\pi}{2}}^{\frac{\pi}{2}}\ln\left[\frac{\sinh(\eta_{+}+ix+\frac{\eta_{+}+i\pi}{2})}{\sinh(\eta_{+}+ix-\frac{\eta_{+}+i\pi}{2})}
 \frac{\sinh(-\eta_{+}+ix+\frac{\eta_{+}+i\pi}{2})}{\sinh(-\eta_{+}+ix-\frac{\eta_{+}+i\pi}{2})}\right]\rho(x)dx \nonumber \\
&&\qquad -\frac{i}{2}\left[\ln\frac{\sinh(ip+\frac{\eta_{+}+i\pi}{2})}{\sinh(ip-\frac{\eta_{+}+i\pi}{2})}+\ln\frac{\sinh(iq+\frac{\eta_{+}+i\pi}{2})}{\sinh(iq-\frac{\eta_{+}+i\pi}{2})}\right].
\end{eqnarray}

The momentum carried by the  elementary excitation is
\begin{eqnarray}\label{mom_ex_im_ex_gap}
		K=\zeta(p)+\zeta(q),
\end{eqnarray}
with
\begin{eqnarray}\label{mom_ex_im_ex_gap_e0} \zeta(t)=\sum_{k=1}^{\infty}(-1)^{k}\frac{\sin(2kt)}{k}e^{-\eta_{+}k}\tanh(\eta_{+}k)-\frac{i}{2}\ln\left[-\frac{\cosh(it+\frac{\eta_{+}}{2})}{\cosh(it-\frac{\eta_{+}}{2})}\right].
\end{eqnarray}
Comparing Eqs.(\ref{energy_ex_im_ex_gap}) and (\ref{mom_ex_im_ex_gap_e0}), we obtain the dispersion relation as shown in Fig.\ref{Dispersion relation 3-image}.
It seems that the dispersion relation takes the same form as that in the periodic boundary condition \cite{Takahashi99}.

\section{Conclusion}\label{6}

In conclusion, an analytic method to derive the root patterns of transfer matrix of quantum integrable models without $U(1)$ symmetry is proposed. It is found that by choosing a proper set of inhomogeneity parameters, the root patterns do not change but only alter the density of distributions. This allows us to derive the density of roots  and to compute the eigenenergy in the thermodynamic limit. This method can be naturally applied to other quantum integrable models.
\section*{Acknowledgments}

The financial supports from the National Natural Science Foundation of China (Grant Nos. 12074410, 12047502, 11934015,
11975183, 11947301 and 11774397), Major Basic Research Program of Natural Science of
Shaanxi Province (Grant Nos. 2017KCT-12 and 2017ZDJC-32), Australian
Research Council (Grant No. DP 190101529), the Strategic
Priority Research Program of the Chinese Academy of Sciences (Grant No. XDB33000000),
the fellowship of China Postdoctoral Science Foundation (Grant No. 2020M680724), and
Double First-Class University Construction Project of
Northwest University are gratefully acknowledged.

\end{document}